\documentclass[10pt]{amsart}

\usepackage[latin1]{inputenc}
\usepackage{amsmath}
\usepackage{amssymb,amsthm}
\usepackage[mathscr]{eucal}

\newcommand\erdos{Erd\H{o}s}

\newtheorem{thm}{Theorem}[section]
\newtheorem{prop}[thm]{Proposition}
\newtheorem{lemma}[thm]{Lemma}
\theoremstyle{definition}
\newtheorem{example}[thm]{Example}
\newtheorem*{remark}{Remark}

\newcommand\stopbox{\hfill\rule{1ex}{1ex}}

\DeclareMathOperator{\curl}{curl}

\let\epsilon=\varepsilon

\newcommand{\gnorm}[1]{\left\lvert\mspace{-1.7mu}\left\lvert\mspace{-1.7mu}\left\lvert#1%
\right\rvert\mspace{-1.7mu}\right\rvert\mspace{-1.7mu}\right\rvert}

\newcommand\vect[1]{\ensuremath{\mathbf{#1}}}

\newcommand\distribD{\ensuremath{\mathcal{D}}}

\newcommand\calD{\ensuremath{\mathscr{D}}}
\newcommand\defD{\ensuremath{\mathscr{D}}}

\newcommand\dz{\ensuremath{\frac{\partial}{\partial z}}}
\newcommand\dzbar{\ensuremath{\frac{\partial}{\partial \bar{z}}}}

\newcommand\diffe{\partial}
\newcommand\diffm{\tilde{\partial}}
\newcommand\dze{\ensuremath{\diffe_z}}

\newcommand\dzbare{\ensuremath{\diffe_{\bar{z}}}}
\newcommand\dzm{\ensuremath{\diffm_z}}

\newcommand\dzbarm{\ensuremath{\diffm_{\bar{z}}}}

\newcommand\pfe{\ensuremath{\pi}}
\newcommand\pfm{\ensuremath{\mathfrak{p}}}
\newcommand\paulienr{\ensuremath{\mathcal{P}}}
\newcommand\paulie{\ensuremath{P}}
\newcommand\paulim{\ensuremath{\mathfrak{P}}}

\title[Zero modes of the Pauli operator]{On the Aharonov-Casher formula for different self-adjoint extensions of the Pauli operator with singular magnetic field}
\author[M. Persson]{Mikael Persson}
\address[M. Persson]{Department of Mathematics \\
                        Chalmers University of Technology \\
                        and University of Gothenburg \\
                        Eklandagatan 86, S-412 96 Gothenburg \\
                        Sweden}
\email{mickep@math.chalmers.se}

\begin{document}

\begin{abstract}
\noindent Two different self-adjoint Pauli extensions describing a spin-1/2 two-dimen\-sional quantum system with singular magnetic field are studied. An Aharonov-Casher type formula is proved for the maximal Pauli extension and it is also checked that this extension can be approximated by operators corresponding to more regular magnetic fields.
\end{abstract}

\maketitle

\section{Introduction}
\label{sec:intro}

\noindent Two-dimensional spin-$1/2$ quantum systems involving magnetic fields are described by the self-adjoint Pauli operator. One interesting question about such systems is the appearance of zero modes (eigenfunctions with eigenvalue zero). Aharonov and Casher proved in~\cite{ac} that if the magnetic field is bounded and compactly supported, then zero modes can arise, and the number of zero modes is simply connected to the total flux of the magnetic field. Since then, Aharonov-Casher type formulas have been proved for more and more singular magnetic fields in different settings, see~\cite{cfks,gegr,lali,mi}. Recently they were proved for measure-valued magnetic fields in~\cite{ervo} by \erdos{} and Vougalter.

We are interested in the Pauli operator when the magnetic field consists of a regular part with compact support and a singular part with a finite number of Aharonov-Bohm (AB) solenoids~\cite{ab}. The Pauli operator for such singular magnetic fields, defined initially on smooth functions with support not touching the singularities, is not essentially self-adjoint. Thus there are several ways of defining the self-adjoint Pauli extension, depending on what boundary conditions one sets at the AB solenoids, see \cite{adte,dast,exstvy,gest,gest2}. Different extensions describe different physics, and there is a discussion going on about which extensions describe the real physical situation. 

There are two possible approaches to making the choice of the extension: trying to describe boundary conditions at the singularities by means of modelling actual interaction of the particle with an AB flux, or considering approximations of singular fields by regular ones, see~\cite{bopu,ta}. We are going to study the maximal extension introduced in~\cite{gegr}, called the Maximal Pauli operator, and compare it with the extension defined in~\cite{ervo}, that we will call the EV Pauli operator. These two extensions were recently studied in~\cite{rosh} in the presence of infinite number of AB solenoids, and it was proved that a magnetic field with infinite flux gives an infinite-dimensional space of zero modes for both extensions.

When studying the Pauli operator in the presence of AB solenoids one must always keep in mind the possibility to reduce the intensities of solenoids by arbitrary integers by means of singular gauge transformations. In Section~\ref{sec:def} we define both extensions via quadratic forms. The Maximal Pauli operator can be defined directly for arbitrary strength of the AB fluxes, while the EV Pauli operator has to be defined via gauge transformations if the AB intensities do not belong to the interval $(-1,1)$. The EV Pauli operator is not gauge invariant. However, following \cite{ervo}, we always make a reduction of the AB intensities to the interval $[-1/2,1/2)$. Hence the EV Pauli operator is not uniquely defined for AB intensities belonging to $(-1,1)\setminus [-1/2,1/2)$, see Section~\ref{sec:def}. Moreover, the asymmetry of the interval $[-1/2,1/2)$, leads to the absence of the invariance of spectral properties of the EV Pauli operator under the changing sign of the magnetic field, the latter property being natural to expect.

For the Dirac operators with strongly singular magnetic field the question on the number of zero modes was considered in \cite{hiog}. The definition of the self-adjoint operator considered there is close to the one in Erdös-Vougalter, however it is not gauge invariant, therefore the Aharonov Casher-type formula obtained in \cite{hiog} depends on intensity of each AB solenoid separtely.

In Section~\ref{sec:prop} we establish that the Maximal Pauli operator is gauge invariant and that changing the sign of the magnetic field leads to anti-unitarily equivalence. Our main result is the Aharonov-Casher type formula for the Maximal Pauli operator. An interesting fact is that this operator can have both spin-up and spin-down zero modes, in contrary to the EV Pauli operator and the Pauli operator for less singular magnetic fields, which have either spin-up or spin-down zero modes, but not both. In~\cite{gegr} a setting with an infinite lattice of AB solenoids with equal AB flux at each solenoid is studied, having both spin-up and spin-down zero modes, both with infinite multiplicity.

In Section~\ref{sec:approx} we discuss the approximation by more regular fields in the sense of Borg and Pulé, see~\cite{bopu}. It turns out that the Maximal Pauli operator can and the EV Pauli operator can not be approximated in this way. However, different ways of approximating the magnetic field may lead to different results, see~\cite{bovo,ta}.

\section{Definition of the Pauli operators}
\label{sec:def}

\noindent The Pauli operator is formally defined as

\begin{displaymath}
P=\left(\sigma\cdot\left(-i\nabla+\vect{A}\right)\right)^2=\left(-i\nabla+\vect{A}\right)^2+\sigma_3 B
\end{displaymath}
on $L_2(\mathbb{R}^2)\otimes \mathbb{C}^2$. Here $\sigma=(\sigma_1,\sigma_2)$, where $\sigma_1$, $\sigma_2$ and $\sigma_3$ are the Pauli matrices

\begin{displaymath}
\sigma_1=
\begin{pmatrix}
0 & 1\\
1 & 0
\end{pmatrix}
,\quad
\sigma_2=
\begin{pmatrix}
0 & -i\\
i & 0
\end{pmatrix}
,\ \text{and}\quad
\sigma_3=
\begin{pmatrix}
1 & 0\\
0 & -1
\end{pmatrix},
\end{displaymath}
$\vect{A}$ is the real magnetic vector potential and $B=\curl(\vect{A})$ is the magnetic field. This definition does not work if the magnetic field $B$ is too singular, see the discussion in~\cite{ervo,so}. If $\vect{A}\in L_{2,\text{loc}}(\mathbb{R}^2)$, using the notations $\Pi_k = -i\partial_k+A_k$, for $k=1,2$, $Q_\pm= \Pi_1\pm i\Pi_2$ and $\lambda$ for the Lebesgue measure, the Pauli operator can be defined via the quadratic form
\begin{equation}
\label{eq:kvadform}
p[\psi]=\|Q_+\psi_+\|^2+\|Q_-\psi_-\|^2=\int|\sigma\cdot(-i\nabla+\vect{A})\psi|^2d\lambda(x),
\end{equation}
the domain being the closure in the sense of the metrics $p[\psi]$ of the core consisting of smooth compactly supported functions. With this notation, we can write the Pauli operator $P$ as

\begin{equation}
\label{eq:paulinotion}
P=\begin{pmatrix} P_+ & 0\\ 0 & P_-\end{pmatrix}=\begin{pmatrix} Q_+^*Q_+ & 0\\ 0 & Q_-^*Q_-\end{pmatrix}.
\end{equation}

However, defining the Pauli operator via the quadratic form $p[\psi]$ in~\eqref{eq:kvadform} requires that the vector potential $\vect{A}$ belongs to $L_{2,\text{loc}}(\mathbb{R}^2)$, otherwise $p[\psi]$ can be infinite for nice functions $\psi$, see~\cite{so}. If the magnetic field consists of only one AB solenoid located at the origin with intensity (flux divided by $2\pi$) $\alpha$, then the magnetic vector potential $\vect{A}$ is given by $\vect{A}(x_1,x_2)=\frac{\alpha}{x_1^2+x_2^2}(-x_2,x_1)$ which is not in $L_{2,\text{loc}}(\mathbb{R}^2)$. Here, and elsewhere we identify a point $(x_1,x_2)$ in the two-dimensional space $\mathbb{R}^2$ with $z=x_1+ix_2$ in the complex plan $\mathbb{C}$.

Following~\cite{ervo}, we will define the Pauli operator via another quadratic form, that agrees with $p[\psi]$ for less singular magnetic fields. We start by describing the magnetic field.

Even though the Pauli operator can be defined for more general magnetic fields, in order to demonstrate the main features of the study, without extra technicalities, we restrict ourself to a magnetic field consisting of a sum of two parts, the first being a smooth function with compact support, the second consisting of finitely many AB solenoids. Let $\Lambda=\{z_j\}_{j=1}^n$ be a set of distinct points in $\mathbb{C}$ and let $\alpha_j\in\mathbb{R}\setminus\mathbb{Z}$. The magnetic field we will study in this paper has the form
\begin{equation}
\label{eq:magnet}
B(z)=B_0(z)+\sum_{j=1}^n 2\pi\alpha_j\delta_{z_j},
\end{equation}
where $B_0\in C_0^1(\mathbb{R}^2)$. In \cite{ervo} the magnetic field is given by a signed real regular Borel measure $\mu$ on $\mathbb{R}^2$ with locally finite total variation. It is clear that $\mu=B_0(z)d\lambda(z)+\sum_{j=1}^n 2\pi\alpha_j\delta_{z_j}$ is such a measure.

The function $h_0$ given by
\begin{displaymath}
h_0(z)=\frac{1}{2\pi}\int \log|z-z'|B_0(z')d\lambda(z')
\end{displaymath}
satisfies $\Delta h_0=B_0$ since $B_0\in C_0^1(\mathbb{R}^2)$ and $\Delta \log|z-z_j|=2\pi\delta_{z_j}$ in the sense of distributions. The function
\begin{displaymath}
h(z)=h_0(z)+\sum_{j=1}^n \alpha_j\log|z-z_j|
\end{displaymath}
satisfies $\Delta h=B$. It is easily seen that $h_0(z)\sim\Phi_0\log|z|$ as $|z|\to\infty$, and thus the asymptotics of $e^{h(z)}$ is
\begin{displaymath}
e^{\pm h(z)}\sim
\begin{cases}
|z|^{\pm \Phi}, & |z|\to\infty\\
|z-z_j|^{\pm \alpha_j}, & z\to z_j,
\end{cases}
\end{displaymath}
where $\Phi_0=\frac{1}{2\pi}\int B_0(z)d\lambda(z)$ and $\Phi=\frac{1}{2\pi}\int B(z)d\lambda(z)=\Phi_0+\sum_{j=1}^n\alpha_j$.

We are now ready to define the two self-adjoint Pauli operators. The decisive difference between them is the sense in which we are taking derivatives. This leads to different domains, and, as we will see in later sections, to different properties of the operators. Let us introduce notations for taking derivatives on the different spaces of distributions. Remember that $\Lambda=\{z_j\}_{j=1}^n$ is a finite set of distinct points in $\mathbb{C}$. We let the derivatives in $\distribD'(\mathbb{R}^2)$ be denoted by $\diffe$ and the derivatives in $\distribD'(\mathbb{R}^2\setminus\Lambda)$ be denoted by $\diffe$ with a tilde over it, that is $\diffm$. Thus, for example, by $\dze$ we mean $\dz$ in the space $\distribD'(\mathbb{R}^2)$ and by $\dzm$ we mean $\dz$ in the space $\distribD'(\mathbb{R}^2\setminus\Lambda)$.

\subsection{The EV Pauli operator} 
We follow~\cite{ervo} and define the sesquilinear forms $\pfe_+$ and $\pfe_-$ by
\begin{align*}
\pfe_+^h(\psi_+,\xi_+)&=4\int \overline{\dzbare\left(e^{-h}\psi_+\right)}\dzbare\left(e^{-h}\xi_+\right)e^{2h}d\lambda(z),\\
\defD(\pfe^h_+)&=\left\{\psi_+\in L_2(\mathbb{R}^2)\ :\ \pfe^h_+(\psi_+,\psi_+)<\infty\right\},
\end{align*}
and
\begin{align*}
\pfe_-^h(\psi_-,\xi_-)&=4\int \overline{\dze\left(e^{h}\psi_-\right)}\dze\left(e^{h}\xi_-\right)e^{-2h}d\lambda(z),\\
\defD(\pfe^h_-)&=\left\{\psi_-\in L_2(\mathbb{R}^2)\ :\ \pfe^h_-(\psi_-,\psi_-)<\infty\right\}.
\end{align*}
Now set
\begin{align*}
\pfe^h(\psi,\xi)&=\pfe_+^h(\psi_+,\xi_+)+\pfe_-^h(\psi_-,\xi_-),\\
\defD(\pfe^h)&=\defD(\pfe^h_+)\oplus\defD(\pfe^h_-)=\big\{\psi=\begin{pmatrix}\psi_+\\ \psi_-\end{pmatrix}\in L_2(\mathbb{R}^2)\otimes\mathbb{C}^2\ :\ \pi^h(\psi,\psi)<\infty\big\}.
\end{align*}
Let us make more accurate the description of the domains of the forms $\pfe_\pm^h$ and $\pfe^h$. For example, what is required of a function $\psi_+$ to be in $\defD(\pfe^h_+)$? It should belong to $L_2(\mathbb{R}^2)$, and the expression
\begin{displaymath}
\pfe_+^h(\psi_+,\psi_+)=4\int \left|\dzbare\left(e^{-h}\psi_+\right)\right|^2e^{2h}d\lambda(z)
\end{displaymath}
should have a meaning and be finite. This means that the distribution $\dzbare\left(e^{-h}\psi_+\right)$ actually must be a function and its modulus multiplied with $e^h$ must belong to $L_2(\mathbb{R}^2)$, that is $|\dzbare\left(e^{-h}\psi_+\right)|e^h\in L_2(\mathbb{R}^2)$. This forces all the intensities $\alpha_j$ to be in the interval $(-1,1)$, see~\cite{ervo}.

Next we define the norm by

\begin{displaymath}
\gnorm{\psi}^2_{\pfe^h}=\gnorm{\psi_+}^2_{\pfe_+^h}+\gnorm{\psi_-}^2_{\pfe_-^h},
\end{displaymath}
where
\begin{displaymath}
\gnorm{\psi_+}^2_{\pfe_+^h}=\|\psi_+\|^2+\left\|\dzbare\left(e^{-h}\psi_+\right)e^h\right\|^2
\end{displaymath}
and
\begin{displaymath}
\gnorm{\psi_-}^2_{\pfe_-^h}=\|\psi_-\|^2+\left\|\dze\left(e^{h}\psi_-\right)e^{-h}\right\|^2.
\end{displaymath}
This form $\pi^h$ is symmetric, nonnegative and closed with respect to $\|\cdot\|$, again see~\cite{ervo}, and hence it defines a unique self-adjoint operator $\paulienr_h$ via

\begin{equation}
\label{eq:pdefenr}
\defD(\paulienr_h)=\{\psi\in\defD(\pfe^h)\ :\ \pfe^h(\psi,\cdot)\in \left(L_2(\mathbb{R}^2)\otimes\mathbb{C}^2\right)\}
\end{equation}
and
\begin{equation}
\label{eq:pdomenr}
(\paulienr_h\psi,\xi)=\pfe^h(\psi,\xi),\quad \psi\in\defD(\paulienr_h),\xi\in\defD(\pfe^h).
\end{equation}
We call this operator $\paulienr_h$ the \emph{non-reduced EV Pauli operator}.

If some intensities $\alpha_j$ belongs to $\mathbb{R}\setminus[-1/2,1/2)$, we let $\alpha_j^*$ be the unique real number in $[-1/2,1/2)$ such that $\alpha_j$ and $\alpha_j^*$ differ only by an integer, that is $\alpha_j^*-\alpha_j=m_j\in\mathbb{Z}$. We define the \emph{reduced EV Pauli operator} (or just the \emph{EV Pauli operator}), $\paulie_h$, to be
\begin{equation}
\label{eq:pdefe}
\paulie_h=\exp(i\phi)\paulienr_h\exp(-i\phi)
\end{equation}
where $\phi(z)=\sum_{j=1}^n m_j \arg(z-z_j)$. Hence, if there are some $\alpha_j$ outside the interval $(-1,1)$ only the reduced EV Pauli operator is well-defined. If all the intensities $\alpha_j$ belong to the interval $[-1/2,1/2)$ then we do not have to perform the reduction and hence there is only one definition. However, if there are intensities $\alpha_j$ inside the interval $(-1,1)$ but outside the interval $[-1/2,1/2)$ then we have two different definitions of the EV Pauli operator, the direct one and the one obtained by reduction. In the next section we will show that these two operators are not the same.

\subsection{The Maximal Pauli operator}
Now, again, let $\alpha_j\in\mathbb{R}\setminus\mathbb{Z}$. We define the forms
\begin{align*}
\pfm_+^h(\psi_+,\xi_+)&=4\int \overline{\dzbarm\left(e^{-h}\psi_+\right)}\dzbarm\left(e^{-h}\xi_+\right)e^{2h}d\lambda(z),\\
\defD(\pfm^h_+)&=\left\{\psi_+\in L_2(\mathbb{R}^2)\ :\ \pfm^h_+(\psi_+,\psi_+)<\infty\right\},
\end{align*}
and
\begin{align*}
\pfm_-^h(\psi_-,\xi_-)&=4\int \overline{\dzm\left(e^{h}\psi_-\right)}\dzm\left(e^{h}\xi_-\right)e^{-2h}d\lambda(z),\\
\defD(\pfm^h_-)&=\left\{\psi_-\in L_2(\mathbb{R}^2)\ :\ \pfm^h_-(\psi_-,\psi_-)<\infty\right\}.
\end{align*}
Now set
\begin{align*}
\pfm^h(\psi,\xi)&=\pfm_+^h(\psi_+,\xi_+)+\pfm_-^h(\psi_-,\xi_-),\\
\defD(\pfm^h)&=\defD(\pfm^h_+)\oplus\defD(\pfm^h_-)=\big\{\psi=\begin{pmatrix}\psi_+\\ \psi_-\end{pmatrix}\in L_2(\mathbb{R}^2)\otimes\mathbb{C}^2\ :\ \pfm^h(\psi,\psi)<\infty\big\}.
\end{align*}
Again, let us make clear about the domains of the forms. For a function $\psi_+$ to be in $\defD(\pfm_+^h)$ it is required that $\psi_+\in L_2(\mathbb{R}^2)$ and that $\dzm(e^{-h}\psi_+)$ is a function. After taking the modulus of this derivative and multiplying by $e^h$ we should get into $L_2(\mathbb{R}^2\setminus\Lambda)$, that is $|\dzbarm(e^{-h}\psi_+)|e^h\in L_2(\mathbb{R}^2\setminus\Lambda)$. Note that the form $\pfm^h$ does not feel the AB fluxes at $\Lambda$ since the derivatives are taken in the space $\distribD'(\mathbb{R}^2\setminus\Lambda)$, and integration does not feel $\Lambda$ either since $\Lambda$ has Lebesgue measure zero. This enable the AB solenoids to have intensities that lies outside $(-1,1)$.

Also, define the norm

\begin{displaymath}
\gnorm{\psi_h}^2_{\pfm^h}=\gnorm{\psi_+}^2_{\pfm_+^h}+\gnorm{\psi_-}^2_{\pfm_-^h},
\end{displaymath}
where
\begin{displaymath}
\gnorm{\psi_+}^2_{\pfm_+^h}=\|\psi_+\|^2+\left|\left|\dzbarm\left(e^{-h}\psi_+\right)e^h\right|\right|^2
\end{displaymath}
and
\begin{displaymath}
\gnorm{\psi_-}^2_{\pfm_-^h}=\|\psi_-\|^2+\left|\left|\dzm\left(e^{h}\psi_-\right)e^{-h}\right|\right|^2.
\end{displaymath}

\begin{prop} The form $\pfm^h$ defined above is symmetric, nonnegative and closed with respect to $\|\cdot\|$.
\end{prop}

\begin{proof} It is clear that $\pfm^h$ is symmetric and nonnegative. Let $\psi_n=(\psi_{n,+},\psi_{n,-})$ be a Cauchy sequence in the norm $\gnorm{\cdot}_{\pfm^h}$. This implies that $\psi_{n,\pm}\to\psi_{\pm}$ in $L_{2}(d\lambda(z))$, $\dzbarm\left(e^{-h}\psi_{n,+}\right)\to u_+$ in $L_2(e^{2h}d\lambda(z))$ and $\dzm(e^h\psi_{n,-})\to u_-$ in $L_2(e^{-2h}d\lambda(z))$. We have to show that $\dzbarm\left(e^{-h}\psi_{+}\right)= u_+$ and $\dzm(e^h\psi_{-})=u_-$. For any test-function $\phi\in C_0^\infty(\mathbb{R}^2\setminus\Lambda)$,

\begin{align*}
\left|\int\bar{\phi}\left(u_+-\dzbarm\left(e^{-h}\psi_{+}\right)\right)d\lambda(z)\right|& \leq \left|\int \bar{\phi}\left(u_+-\dzbarm\left(e^{-h}\psi_{n,+}\right)\right)\right|\\
\mbox{}&\mbox{}\quad +\left|\int\dzbarm(\bar\phi)e^{-h}\left(\psi_+-\psi_{n,+}\right)\right|\\
\mbox{}& \leq \|\bar\phi e^{-h}\|\cdot \left\|u_+-\dzbarm\left(e^{-h}\psi_{n,+}\right)\right\|_{L_2(e^{2h})}\\
\mbox{}&\mbox{}\quad +\left\|\dzbarm(\bar\phi)e^{-h}\right\|\cdot \|\psi_+-\psi_{n,+}\|.
\end{align*}
The last expression tends to zero as $n\to\infty$, since the first terms in each sum is bounded (thanks to $\phi$) and the other one tends to zero. The proof is the same for the spin down component. This shows that $\pfm^h$ is closed.
\end{proof}

\noindent Hence $\pfm^h$ defines a unique self-adjoint operator $\paulim_h$ via
\begin{equation}
\label{eq:paulidef}
\calD(\paulim_h)=\{\psi\in\calD(\pfm^h)\ :\ \pfm^h(\psi,\cdot)\in \left(L_2(\mathbb{R}^2)\otimes\mathbb{C}^2\right)\}
\end{equation}
and
\begin{equation}
\label{eq:paulidom}
(\paulim_h\psi,\xi)=\pfm^h(\psi,\xi),\quad \psi\in\calD(\paulim_h),\xi\in\calD(\pfm^h).
\end{equation}
We call this operator $\paulim_h$ the \emph{Maximal Pauli operator}.

\section{Properties of the Pauli operators}
\label{sec:prop}

In this section we will compare some properties of the two Pauli operators $\paulie_h$ and $\paulim_h$ defined in the previous section. We start by showing that $\paulim_h$ is gauge invariant while $\paulie_h$ is not.

\subsection{Gauge transformations}

Let $B(z)=B_0(z)+\sum_{j=1}^n2\pi \alpha_j\delta_{z_j}$ be the same magnetic field as before and let $\hat{B}(z)$ be another magnetic field that differs from $B(z)$ only by some multiples of the delta functions, that is $\hat{B}(z)-B(z)=\sum_{j=1}^n 2\pi m_j\delta_{z_j}$, where $m_j$ are integers, not all zero. Then the corresponding scalar potentials $\hat{h}(z)$ and $h(z)$ differ only by the corresponding logarithms $\hat{h}(z)-h(z)=\sum_{j=1}^nm_j\log|z-z_j|$. Now with $\phi(z)=\sum_{j=1}^nm_j\arg(z-z_j)$ we get $\hat{h}(z)+i\phi(z)=h(z)+\sum_{j=1}^nm_j\log(z-z_j)$. This function is multivalued, however, since $m_j$ are integers, we have
\begin{align}
\label{eq:diffe}
\dzbare\left(\hat{h}(z)+i\phi(z)\right)&=\dzbare h(z)+\sum_{j=1}^nm_j\dzbare\log(z-z_j),\\
\label{eq:diffm}
\dzbarm\left(\hat{h}(z)+i\phi(z)\right)&=\dzbarm h(z),\ \text{and}\\
\label{eq:exprule}
e^{\hat{h}+i\phi}&=e^h\prod_{j=1}^m(z-z_j)^{m_j}.
\end{align}
To see that $\paulie_h$ is not gauge invariant it is enough to look at an example. Let $n=1$, $z_1=0$, $\alpha_1=-1/2$ and $m_1=1$, so the two magnetic fields are $B(z)=B_0(z)-\pi\delta_0$ and $\hat{B}(z)=B_0(z)+\pi\delta_0$. 
The scalar potentials are given by $h(z)=h_0(z)-\frac12\log|z|$ and $\hat{h}(z)=h_0(z)+\frac12\log|z|$ respectively, where $h_0(z)$ is a smooth function with asymptotics $\Phi_0\log|z|$ as $|z|\to\infty$. We should show that $\defD(\pfe^{\hat{h}})$ is not given by $e^{-i\phi}\defD(\pfe^h)$, where $\phi(z)=\arg(z)$. Then it follows that $\pfe^h$ and $\pfe^{\hat{h}}$ do not define unitarily equivalent operators.

Let $\psi_+\in\defD(\pfe_+^h)$. This means, in particular, that $\dzbare(\psi_+e^{-h})$ belongs to $L_{1,\text{loc}}(\mathbb{R}^2)$. Now let $\hat{\psi}_+=e^{-i\phi}\psi_+$. Then, according to~\eqref{eq:exprule} we get
\begin{displaymath}
\dzbare(\hat{\psi}_+e^{-\hat{h}})=\dzbare(\psi_+e^{-\hat{h}-i\phi})=\dzbare\left(\frac{\psi_+e^{-h}}{z}\right)=\dzbare(\psi_+e^{-h})\frac1z+\psi_+e^{-h}\pi\delta_0
\end{displaymath}
which is not in $L_{1,\text{loc}}(\mathbb{R}^2)$ since it is a distribution involving $\delta_0$ (for non-smooth $\psi_+$ it is not even well-defined). Thus we have $\defD(\pfe_+^{\hat{h}})\neq e^{-i\phi}\defD(\pfe_+^h)$ and hence $\defD(\pfe^{\hat{h}})\neq e^{-i\phi}\defD(\pfe^h)$ so $\pfe^h$ and $\pfe^{\hat{h}}$ are not defining unitarily equivalent operators.

Let us now study what happens with $\pfm^h$ when we do gauge transforms. Let $\psi=(\psi_+,\psi_-)^t\in\defD(\pfm^h)$. We should check that $e^{-i\phi}\psi$ belongs to $\defD(\pfm^{\hat{h}})$, where $\phi(z)=\sum_{j=1}^n m_j\arg(z-z_j)$ is the harmonic conjugate to $\hat{h}(z)-h(z)$. We do this for $\pfm_+^{\hat{h}}$. It is similar for $\pfm_-^{\hat{h}}$. Since $\psi_+\in\defD(\pfm_+^h)$ we know that $\dzbarm(\psi_+e^{-h})\in L_{1,\text{loc}}(\mathbb{R}^2\setminus\Lambda)$. Let us check that $\dzbarm(\hat{\psi}_+e^{-\hat{h}})\in L_{1,\text{loc}}(\mathbb{R}^2\setminus\Lambda)$. Again, by~\eqref{eq:exprule} we have
\begin{align*}
\dzbarm(\hat{\psi}_+e^{-\hat{h}})&=\dzbarm\left(\psi_+e^{-h}\prod_{j=1}^n(z-z_j)^{-m_j}\right)\\
&=\dzbarm\left(\psi_+e^{-h}\right)\prod_{j=1}^n(z-z_j)^{-m_j}+\psi_+e^{-h}\dzbarm\left(\prod_{j=1}^n(z-z_j)^{-m_j}\right),
\end{align*}
which clearly belongs to $L_{1,\text{loc}}(\mathbb{R}^2\setminus\Lambda)$.

Next we should check that $|\dzbarm(\hat{\psi}_+e^{-\hat{h}})|e^{\hat{h}}$ belongs to $L_2(\mathbb{R}^2\setminus\Lambda)$ under the assumption that $|\dzbarm(\psi_+e^{-h})|e^{h}$ belongs to $L_2(\mathbb{R}^2\setminus\Lambda)$. A calculation using~\eqref{eq:diffm} and~\eqref{eq:exprule} gives

\begin{small}
\begin{align}
\nonumber\left|\dzbarm\left(e^{-\hat{h}}\hat{\psi}_+\right)\right|e^{\hat{h}} &= \left|\dzbarm\left(e^{-\hat{h}-i\phi}\psi_+(z)\right)\right|e^{\hat{h}}\\
\label{eq:ltwo}
&= \left|\left(\dzbarm(-h(z))\psi_++\dzbarm\psi_+(z)\right)e^{-h}\prod_{j=1}^n(z-z_j)^{-m_j}\right|e^{h}\prod_{j=1}^n|z-z_j|^{m_j}\\
\nonumber&=\left|\dzbarm\left(e^{-h}\psi_+\right)\right|e^{h}.
\end{align}
\end{small}%
Hence $\psi_+\in\defD(\pfm_+^h)$ implies $\hat{\psi}_+=e^{-i\phi}\psi_+\in\defD(\pfm_+^{\hat{h}})$. In a similar way it follows that $\psi_-\in\defD(\pfm_-^h)$ implies that $\hat{\psi}_-=e^{-i\phi}\psi_-\in\defD(\pfm_-^{\hat{h}})$. Thus $e^{-i\phi}\defD(\pfm^h)\subset\defD(\pfm^{\hat{h}})$. In the same way we can show that $e^{i\phi}\defD(\pfm^{\hat{h}})\subset\defD(\pfm^h)$, and thus we can conclude that $e^{-i\phi}\defD(\pfm^h)=\defD(\pfm^{\hat{h}})$. From the calculation in~\eqref{eq:ltwo} and a similar calculation for $\psi_-$ it also follows that 

\begin{align*}
\pfm^{\hat{h}}\left(e^{-i\phi}\psi,e^{-i\phi}\psi\right) & = 4\int\left|\dzbarm\left(e^{-\hat{h}-i\phi}\psi_+\right)\right|^2e^{2\hat{h}}+\left|\dzm\left(e^{\hat{h}-i\phi}\psi_-\right)\right|^2e^{-2\hat{h}}d\lambda(z)\\
& = 4\int\left|\dzbarm\left(e^{-h}\psi_+\right)\right|^2e^{2h}+\left|\dzm\left(e^{h}\psi_-\right)\right|^2e^{-2h}d\lambda(z)\\
& = \pfm^h(\psi,\psi).
\end{align*}
Hence we can conclude that if $\psi\in\defD(\paulim_h)$ and $\xi\in\defD(\pfm^h)$ then $e^{-i\phi}\psi\in\defD(\paulim_{\hat{h}})$ and $e^{-i\phi}\xi\in\defD(\pfm^{\hat{h}})$. If we denote by $U_\phi$ the unitary operator of multiplication by $e^{i\phi}$, then we get
\begin{displaymath}
(\paulim_h\psi,\xi)=\pfm^h(\psi,\xi)=\pfm^{\hat{h}}(U_\phi^*\psi,U_\phi^*\xi)=(\paulim_{\hat{h}}U_\phi^*\psi,U_\phi^*\xi)=(U_\phi \paulim_{\hat{h}}U_\phi^*\psi,\xi),
\end{displaymath}
and hence $\paulim_h$ and $\paulim_{\hat{h}}$ are unitarily equivalent. We have proved the following proposition.

\begin{prop}
Let $B$ and $\hat{B}$ be two singular magnetic fields as in \eqref{eq:magnet}, with difference $\hat{B}-B=\sum_{j=1}^n 2\pi m_j \delta_{z_j}$, where $m_j$ are integers, not all equal to zero. Then their corresponding Maximal Pauli operators defined by~\eqref{eq:paulidef} and~\eqref{eq:paulidom} are unitarily equivalent.
\end{prop}

\subsection{Zero modes}
\label{sec:zero}

When studying spectral properties of the operator $\paulim_h$ it is sufficient to consider AB intensities $\alpha_j$ that belong to the interval $(0,1)$, since the operator is gauge invariant. See the discussion after the proof of Theorem~\ref{thm:AC} for more details about what happens when we do gauge transformations.

\begin{lemma}
\label{lemma:asymptot}
Let $c_j\in\mathbb{C}$ and $z_j\in\mathbb{C}$, $j=1,\ldots,n$, where $z_j\neq z_i$ if $j\neq i$ and not all $c_j$ are equal to zero. Then
\begin{equation}
  \label{eq:asympt}
  \sum_{j=1}^n \frac{c_j}{z-z_j}\sim |z|^{-l-1},\quad |z|\to\infty,
\end{equation}
where $l$ is the smallest nonnegative integer such that $\sum_{j=1}^n c_j z_j^l\neq 0$.
\end{lemma}

\begin{proof}
If $|z|$ is large in comparison with all $|z_j|$ we have
\begin{align*}
  \sum_{j=1}^n \frac{c_j}{z-z_j} &= \frac{1}{z}\sum_{j=1}^n \frac{c_j}{1-z_j/z}\\
&= \sum_{k=0}^\infty \left(\sum_{j=1}^n c_jz_j^k\right)\frac{1}{z^{k+1}}\\
&= \left(\sum_{j=1}^n c_jz_j^l\right)\frac{1}{z^{l+1}}+O(|z|^{-l-2})
\end{align*}
and thus $\sum_{j=1}^n \frac{c_j}{z-z_j}\sim |z|^{-l-1}$ as $|z|\to\infty$.
\end{proof}

\begin{remark}We note that $l$ in Lemma~\ref{lemma:asymptot} may never be greater than $n-1$. Indeed, if $l\geq n$ then we would have the linear system of equations $\{\sum_{j=1}^nc_jz_j^k=0\}_{k=0}^{n-1}$. But the determinant of this system is $\prod_{i>j}(z_i-z_j)\neq 0$, and this would force all $c_j$ to be zero.

Note also that for $l\leq n$ we have a system of $l$ equations $\{\sum_{j=1}^nc_jz_j^k=0\}_{k=0}^{l-1}$ with $n$ unknowns $c_j$, and that the $l\times n$ matrix $\{z_j^k\}$ has rank $l$.\stopbox
\end{remark}

\begin{thm}
\label{thm:AC}
Let $B(z)$ be the magnetic field~\eqref{eq:magnet} with all $\alpha_j\in(0,1)$, and let $\paulim_h$ be the Pauli operator defined by~\eqref{eq:paulidef} and \eqref{eq:paulidom} in Section~\ref{sec:def} corresponding to $B(z)$. Then
\begin{equation}
\dim\ker \paulim_h=
\left\{n-\Phi\right\}+\left\{\Phi\right\},
\end{equation}
where $\Phi=\frac{1}{2\pi}\int B(z)d\lambda(z)$, and $\{x\}$ denotes the largest integer strictly less than $x$ if $x>1$ and $0$ if $x\leq 1$. Using the notations $Q_\pm$ introduced in Section~\ref{sec:def}, we also have
\begin{equation}
\dim\ker Q_+=\left\{n-\Phi\right\}\quad \text{and}\quad \dim\ker Q_-=\left\{\Phi\right\}.
\end{equation}
\end{thm}

\begin{proof}
We follow the reasoning originating in~\cite{ac}, with necessary modifications. First we note that $(\psi_+,\psi_-)^t$ belongs to $\ker \paulim_h$ if and only if $\psi_+$ belongs to $\ker Q_+$ and $\psi_-$ belongs to $\ker Q_-$, which is equivalent to
\begin{equation}
\dzbarm\left(e^{-h}\psi_+\right)=0\quad \text{and}\quad \dzm\left(e^{h}\psi_-\right)=0.
\end{equation}
This means exactly that $f_+(z)=e^{-h}\psi_+(z)$ is holomorphic and $f_-(z)=e^h\psi_-(z)$ is antiholomorphic in $z\in\mathbb{R}^2\setminus\Lambda$. It is the change in the domain where the functions are holomorphic that influences the result.

Let us start with the spin-up component $\psi_+$. The function $f_+$ is allowed to have poles of order at most one at $z_j$, $j=1,\ldots,n$, and no others, since $e^h\sim |z-z_j|^{\alpha_j}$ as $z\to z_j$ and $\psi_+=f_+e^h$ should belong to $L_2(\mathbb{R}^2)$. Hence there exist constants $c_j$ such that the function $f_+(z)-\sum_{j=1}^n\frac{c_j}{z-z_j}$ is entire. From the asymptotics $e^h\sim |z|^{\Phi}$, $|z|\to\infty$, it follows that $f_+-\sum_{j=1}^n\frac{c_j}{z-z_j}$ may only be a polynomial of degree at most $N=-\Phi-2$. Hence 
\begin{displaymath}
f_+(z)=\sum_{j=1}^n\frac{c_j}{z-z_j}+a_0+a_1z+\ldots a_Nz^N,
\end{displaymath}
where we let the polynomial part disappear if $N<0$. Now, the asymptotics for $\psi_+$ is
\begin{displaymath}
\psi_+(z)\sim |z|^{-l-1+\Phi}+|z|^{N+\Phi},\quad |z|\to\infty,
\end{displaymath}
where $l$ is the smallest nonnegative integer such that $\sum_{j=1}^n c_jz_j^l\neq0$. To have $\psi_+$ in $L_2(\mathbb{R}^2)$ we take $l$ to be the smallest nonnegative integer strictly greater than $\Phi$. Remember also from the remark after Lemma~\ref{lemma:asymptot} that $l\leq n-1$. We get three cases. If $\Phi<-1$, then all complex numbers $c_j$ can be chosen freely, and a polynomial of degree $\{-\Phi\}-1$ may be added which results $\{n-\Phi\}$ degrees of freedom. If $-1\leq\Phi<n-1$ we have no contribution from the polynomial, and we now have to chose the coefficients $c_j$ such that $\sum_{j=1}^n c_jz_j^k=0$ for $k=0,1,\ldots,l-1$. The dimension of the null-space of the matrix $\{z_j^k\}$ is $n-l=\{n-\Phi\}$. If $\Phi\geq n-1$ then we must have all coefficients $c_j$ equal to zero and we get no contribution from the polynomial. Hence, in all three cases we have $\{n-\Phi\}$ spin-up zero modes.

Let us now focus on the spin-down component $\psi_-$. The function $f_-$ may not have any singularities, since the asymptotics of $e^{-h}$ is $|z-z_j|^{-\alpha_j}$ as $z\to z_j$. Hence $f_-$ must be entire. Moreover, $f_-$ may grow no faster than a polynomial of degree $\Phi-1$ for $\psi_-$ to be in $L_2(\mathbb{R}^2)$. Thus $f_-$ has to be a polynomial of degree at most $\{\Phi\}-1$, which gives us $\{\Phi\}$ spin-down zero modes.
\end{proof}

\noindent The number of zero modes for $\paulim_h$ and $\paulie_h$ are not the same. The Aharonov-Casher theorem for the EV Pauli operator (Theorem 3.1 in \cite{ervo}) states for the field under consideration:

\begin{thm} 
\label{thm:ACEV}
Let $B(z)$ be as in~\eqref{eq:magnet} and let $\hat{B}(z)$ be the unique magnetic field where all AB intensities $\alpha_j$ are reduced to the interval $[-1/2,1/2)$, that is $\hat{B}(z)=B(z)+\sum_{j=1}^n2\pi m_j\delta_{z_j}$, where $\alpha_j+m_j\in[-1/2,1/2)$. Let $\Phi=\frac{1}{2\pi}\int\hat{B}(z)d\lambda(z)$. Then the dimension of the kernel of the EV Pauli operator $\paulie_h$ is given by $\{|\Phi|\}$. All zero modes belong only to the spin-up or only to the spin-down component (depending on the sign of $\Phi$).
\end{thm}

Below we explain by some concrete examples how the spectral properties of the two Pauli operators $\paulim_h$ and $\paulie_h$ differ.

\begin{example}
\label{ex:gauge}
Since $\paulie_h$ is not gauge invariant we must not expect that the number of zero modes of $\paulie_h$ is invariant under gauge transforms. To see that this property in fact can fail, let us look at the Pauli operators $\paulie_{h_1}$ and $\paulie_{h_2}$ induced by the magnetic fields
\begin{align*}
B_1(z)&=B_0(z)+\pi\delta_0,\ \text{and}\\
B_2(z)&=B_0(z)-\pi\delta_0
\end{align*}
respectively, where $B_0$ has compact support and $\Phi_0=\frac{1}{2\pi}\int B_0(z)d\lambda(z)=\frac{3}{4}$. Then $B_2$ is reduced (that is, its AB intensity belong to $[-1/2,1/2)$) but $B_1$ has to be reduced. Due to Theorem~\ref{thm:ACEV}, the EV Pauli operators $\paulie_{h_1}$ and $\paulie_{h_2}$ corresponding to $B_1$ and $B_2$ have no zero modes. However, a direct computation for the non-reduced EV Pauli operator $\paulienr_{h_1}$ corresponding to $B_1$ shows that it actually has one zero mode. The situation is getting more interesting when we look at the operator that should correspond to $B_3=B_0(z)+3\pi\delta_0$. The AB intensity for $B_3$ is too strong so we have to make a reduction. In~\cite{ervo} the reduction is made to the interval $[-1/2,1/2)$, and we have followed this conventions, but physically there is nothing that says that this is the natural choice. Reducing the AB intensity of $B_3$ to $-1/2$ gives an operator with no zero modes and reducing it to $1/2$ gives an operator with one zero mode.

The Maximal Pauli operators $\paulim_{h_1}$, $\paulim_{h_2}$ and $\paulim_{h_3}$ for these three magnetic fields all have one zero mode. This is easily seen by applying Theorem~\ref{thm:AC} to $\paulim_{h_1}$ and then using the fact that the operators are unitarily equivalent.

However, more understanding is achieved when looking more closely at how the eigenfunctions for these three Maximal Pauli operators look like. Let $h_k$ be the scalar potential for $B_k$, $k=1,2,3$. Then, as we have seen before $h_1(z)=h_0(z)+\frac12\log|z|$, $h_2(z)=h_0(z)-\frac12\log|z|$ and $h_3(z)=h_0(z)+\frac32\log|z|$ where $h_0(z)$ corresponds to $B_0(z)$. Following the reasoning from the proof of Theorem~\ref{thm:AC} we see that the solution space to $\paulim_{h_1}\psi=0$ is spanned by $\psi=(0,e^{-h_1})^t$.

Next, we see what the solutions to $\paulim_{h_2}\psi=0$ look like. Now we have $\Phi_2=\frac{1}{2\pi}\int B_2(z)d\lambda(z)=1/4>0$. Let us begin with the spin-up component $\psi_+$. This time, the holomorphic $f_+=e^{-h_2}\psi_+$ may not have any poles since then $\psi_+$ would not belong to $L_2(\mathbb{R}^2)$, and $f_+(z)=e^{-h_2}\psi_+(z)\to0$ as $|z|\to\infty$, so we must have $f_+\equiv 0$, and thus $\psi_+\equiv 0$. For $\psi_-(z)$ to be in $L_2(\mathbb{R}^2)$ it is possible for $f_-$ to have a pole of order $1$ at the origin. Hence there exist a constant $c$ such that $f_-(z)-c/\bar{z}$ is antiholomorphic in the whole plane. The function $f_-(z)\to0$ as $|z|\to\infty$ since the total intensity $\Phi_2>0$. This implies, by Liouville's theorem, that $f_-(z)\equiv c/\bar{z}$, so the solution space to $\paulim_{h_2}\psi=0$ is spanned by $\psi(z)=(0,e^{-h_2}/\bar{z})$.

Finally, let us determine the solutions to $\paulim_{h_3}\psi=0$. Now $\Phi_3=\frac{1}{2\pi}\int B_3(z)d\lambda(z)=9/4$. Consider the spin-up part $\psi_+$. For $\psi_+$ to be in $L_2(\mathbb{R}^2)$ our function $f_+$ may have a pole of order no more than two at the origin. As before, there exist constants $c_1$ and $c_2$ such that $f_+(z)-c_1/z-c_2/z^2$ is entire and its limit is zero as $|z|\to\infty$, and thus $f_+(z)\equiv c_1/z+c_2/z^2$. Again, both $c_1$ and $c_2$ must vanish for $\psi_+$ to be in $L_2(\mathbb{R}^2)$ (otherwise we would not stay in $L_2$ at infinity). Thus $\psi_+\equiv 0$. On the other hand, the function $f_-$ may not have any poles (these poles would push $\psi_-$ out of $L_2(\mathbb{R}^2)$), so it is antiholomorphic in the whole plane. It also may grow no faster than $|z|^{5/4}$ as $|z|\to\infty$, and thus $f_-$ has to be a first order polynomial in $\bar{z}$, that is $f_-(z)=c_0+c_1\bar{z}$. Moreover for $\psi_-$ to be in $L_2(\mathbb{R}^2)$ it must have a zero of order $1$ at the origin, and thus $f_-(z)=c_1\bar{z}$. We conclude that the solutions to $\paulim_{h_3}\psi=0$ are spanned by $(0,\bar{z}e^{-h_3})^t$.
\stopbox
\end{example}

A natural property one should expect of a reasonably defined Pauli operator is that its spectral properties are invariant under the reversing the direction of the magnetic field: $B\mapsto -B$. The corresponding operators are formally anti-unitary equivalent under the transformation $\psi\mapsto \bar{\psi}$ and interchanging of $\psi_+$ and $\psi_-$.

\begin{example}
\label{ex:BminusB}
The number of zero modes for $\paulie_h$ is not invariant under $B(z)\mapsto-B(z)$, which we should not expect since the interval $[-1/2,1/2)$ is not symmetric. We check this by showing that the number of zero modes are not the same. To see this, let $B(z)=B_0(z)+\pi\delta_0$, where $B_0$ has compact support and $\Phi_0=\frac{1}{2\pi}\int B_0(z)d\lambda(z)=\frac{3}{4}$. Then $B$ has to be reduced since the AB intensity at zero is $1/2\not\in[-1/2,1/2)$. After reduction we get the magnetic field $\hat{B}(z)=B_0(z)-\pi\delta_0$, and we can apply Theorem~\ref{thm:ACEV}. Now let $\hat{\Phi}=\frac{1}{2\pi}\int \hat{B}^*d\lambda(z)=\frac14$. Thus the number of zero modes for $\paulie_h$ is $0$. Now look at the Pauli operator $\paulie_{-h}$ defined by the magnetic field $B_-(z)=-B(z)=-B_0(z)-\pi\delta_0$. This magnetic field is reduced and thus we can apply Theorem~\ref{thm:ACEV} directly. The total intensity is $\Phi_-=\frac{1}{2\pi}\int -B(z)d\lambda(z)=-\frac{5}{4}$, so the number of zero modes for $\paulie_{-h}$ is $1$. If $B$ has several AB fluxes then the difference in the number of zero modes of $\paulie_h$ and $\paulie_{-h}$ can be made arbitrarily large.

Now, let us check that the number of zero modes for $\paulim_h$ is invariant under $B(z)\mapsto -B(z)$. Since it is clear that the number of zero modes is invariant under $z\mapsto \bar{z}$ we look instead at how the Pauli operators change when we do $B(z)\mapsto \hat{B}(z)=-B(\bar{z})$. If we set $\zeta=\bar{z}$ we get $\hat{B}(\zeta)=-B(z)$ and the scalar potentials satisfy $\hat{h}(\zeta)=-h(z)$. Now assume that $\psi=(\psi_+(z),\psi_-(z))^t\in\defD(\pfm^h)$. Then
\begin{small}
\begin{align*}
\mbox{}&\pfm^{h(z)}\left(\begin{pmatrix}\psi_+(z)\\ \psi_-(z)\end{pmatrix},\begin{pmatrix}\psi_+(z)\\ \psi_-(z)\end{pmatrix}\right) = \\
\mbox{}&\quad=4\int \left|\dzbarm(\psi_+(z)e^{-h(z)})\right|^2e^{2h(z)}+\left|\dzm(\psi_-(z)e^{h(z)}\right|^2e^{-2h(z)}d\lambda(z)\\
\mbox{}&\quad=4\int\left|\diffm_\zeta(\psi_+(\bar{\zeta})e^{\hat{h}(\zeta)}\right|^2e^{-2\hat{h}(\zeta)}+\left|\diffm_{\bar{\zeta}}(\psi_-(\bar{\zeta})e^{-\hat{h}(\zeta)}\right|^2e^{2\hat{h}(\zeta)}d\lambda(\zeta)\\
\mbox{}&\quad=\pfm^{\hat{h}(\bar{z})}\left(\begin{pmatrix}\psi_-(z)\\ \psi_+(z)\end{pmatrix},\begin{pmatrix}\psi_-(z)\\ \psi_+(z)\end{pmatrix}\right)
\end{align*}
\end{small}%
Hence we see that $(\psi_+,\psi_-)^t$ belongs to $\defD(\paulim_{h(z)})$ if and only if $(\psi_-,\psi_+)^t$ belongs to $\defD(\paulim_{\hat{h}(\bar{z})})$ and then $\paulim_{\hat{h}(\bar{z})}=\paulim_{h(z)}V$ where $V:L_2(\mathbb{R}^2)\otimes\mathbb{C}^2\to L_2(\mathbb{R}^2)\otimes\mathbb{C}^2$ is the isometric operator given by $V((\psi_+,\psi_-)^t)=(\psi_-,\psi_+)^t$. Hence it is clear that $\paulim_{\hat{h}(\bar{z})}$ and $\paulim_{h(z)}$ have the same number of zero modes.
\stopbox
\end{example}

\begin{example}
In the previous example we saw that the number of zero modes for the Maximal Pauli operators corresponding to $B$ and $-B$ are the same. This can easily be seen directly from the Aharonov-Casher formula in Theorem~\ref{thm:AC}. To be able to apply the theorem to $-B=-B_0-\sum_{j=1}^n 2\pi\alpha_j\delta_j$ we have to do gauge transformations, adding $1$ to all the AB intensities, resulting in $\hat{B}=-B_0+\sum_{j=1}^n 2\pi(1-\alpha_j)\delta_j$. Now according to Theorem~\ref{thm:AC} the number of zero modes of $\paulim_{-h}$ is equal to 
\begin{displaymath}
\dim\ker\paulim_{-h}=\{\hat{\Phi}\}+\{n-\hat{\Phi}\}=\{n-\Phi\}+\{\Phi\}=\dim\ker\paulim_h,
\end{displaymath}
where we have used that $\hat{\Phi}=\frac{1}{2\pi}\int \hat{B}d\lambda(z)=n-\Phi$.\stopbox
\end{example}

\section{Approximation by regular fields}
\label{sec:approx}

\noindent We have mentioned that the different Pauli extensions depend on which boundary conditions are induced at the AB fluxes. Let us now make this more precise. Since the self-adjoint extension only depends on the boundary condition at the AB solenoids it is enough to study the case of one such solenoid and no smooth field. For simplicity, let the solenoid be located at the origin, with intensity $\alpha\in(0,1)$, that is, let the magnetic field be given by $B=2\pi\alpha\delta_0$. We consider self-adjoint extensions of the Pauli operator $P$ that can be written in the form
\begin{displaymath}
P=\begin{pmatrix}P_+ & 0\\ 0 & P_-\end{pmatrix}=
\begin{pmatrix}
Q_+^*Q_+ & 0\\
0 & Q_-^*Q_-,
\end{pmatrix}
\end{displaymath}
with some explicitly chosen closed operators $Q_\pm$. It is exactly such extensions $P$ that can be defined by the quadratic form~\eqref{eq:kvadform}. A function $\psi_+$ belongs to $\defD(P_+)$ if and only if $\psi_+$ belongs to $\defD(Q_+)$ and $Q_+\psi_+$ belongs to $\defD(Q_+^*)$, and similarly for $P_-$.

With each self-adjoint extension $P_\pm=Q_{\pm}^*Q_{\pm}$ one can associate (see~\cite{dast,exstvy,gest,ta}) functionals $c_{-\alpha}^\pm$, $c_{\alpha}^{\pm}$, $c_{\alpha-1}^\pm$ and $c_{1-\alpha}^\pm$, by

\begin{align*}
c_{-\alpha}^{\pm}(\psi_\pm) &= \lim_{r\to 0}r^\alpha\frac{1}{2\pi}\int_{0}^{2\pi}\psi_\pm d\theta,\\
c_{\alpha}^{\pm}(\psi_\pm) &= \lim_{r\to 0}r^{-\alpha}\left(\frac{1}{2\pi}\int_{0}^{2\pi}\psi_\pm d\theta-r^{-\alpha}c_{\alpha}^{\pm}(\psi_\pm)\right),\\
c_{\alpha-1}^{\pm}(\psi_\pm) &= \lim_{r\to 0}r^{1-\alpha}\frac{1}{2\pi}\int_{0}^{2\pi}\psi_\pm e^{i\theta} d\theta,\ \text{and}\\
c_{1-\alpha}^{\pm}(\psi_\pm) &= \lim_{r\to 0}r^{\alpha-1}\left(\frac{1}{2\pi}\int_{0}^{2\pi}\psi_\pm e^{i\theta}d\theta-r^{\alpha-1}c_{1-\alpha}^{\pm}(\psi_\pm)\right).
\end{align*}
such that $\psi_\pm\in\defD(P_\pm)$ if and only if
\begin{equation}
\psi_\pm\sim c_{-\alpha}^\pm r^{-\alpha}+c_{\alpha}^\pm r^{\alpha}+c_{\alpha-1}^\pm r^{\alpha-1}e^{-i\theta}+c_{1-\alpha}^\pm r^{1-\alpha}e^{-i\theta}+O(r^\gamma)
\end{equation}
as $r\to 0$, where $\gamma=\min(1+\alpha,2-\alpha)$ and $z=re^{i\theta}$.

Any two nontrivial independent linear relations between these functionals determine a self-adjoint extension. In order that the operator be rotation-invariant, none of these relations may involve both $\alpha$ and $1-\alpha$ terms simultaneously. Accordingly, the parameters $\nu_0^\pm=c_\alpha^\pm/c_{-\alpha}^\pm$ and $\nu_1^\pm=c_{1-\alpha}^\pm/c_{\alpha-1}^\pm$, with possible values in $(-\infty,\infty]$, are introduced in~\cite{bopu}, and it is proved that the operators $P_\pm$ can be approximated by operators with regularized magnetic fields in the norm resolvent sense if and only if $\nu_0^\pm=\infty$ and $\nu_1^\pm\in(-\infty,\infty)$ or if $\nu_0^\pm\in(-\infty,\infty)$ and $\nu_1^\pm=\infty$. We are now going to check what parameters the Maximal and EV Pauli operators corresponds to.

Generally, for the function $\psi_+$ to be in $\defD(P_+)$, it must belong to $\defD(Q_+)$ and $Q_+\psi_+$ must belong to $\defD(Q_+^*)$. We will find out what is required for a function $g$ to be in $\defD(Q_+^*)$. Take any $\phi_+\in\defD(Q_+)$, then the integration by parts on the domain $\epsilon<|z|$ gives

\begin{align*}
\langle g,Q_+\phi_+\rangle &= \lim_{\epsilon\to0}\int_{|z|>\epsilon}g(z)\overline{\dzbar(e^{-h}\phi_+(z))e^h} d\lambda(z)\\
\mbox{}& = -\lim_{\epsilon\to0}\int_{|z|>\epsilon}\dz(g(z)e^h)e^{-h}\overline{\phi_+(z)}d\lambda(z)\\
\mbox{}&\quad\quad -\lim_{\epsilon\to0}\frac{\epsilon}{2}\int_0^{2\pi}g(\epsilon e^{i\theta})\overline{\phi_+(\epsilon e^{i\theta})}e^{-i\theta}d\theta\\
\mbox{}& = \langle -Q_-g,\phi_+\rangle-\lim_{\epsilon\to0}\frac{\epsilon}{2}\int_0^{2\pi}g(\epsilon e^{i\theta})\overline{\phi_+(\epsilon e^{i\theta})}e^{-i\theta}d\theta
\end{align*}
Hence, for $g$ to belong to $\defD(Q_+^*)$ it is necessary and sufficient that

\begin{displaymath}
\lim_{\epsilon\to0}\epsilon\int_0^{2\pi}g(\epsilon e^{i\theta})\overline{\phi_+(\epsilon e^{i\theta})}e^{-i\theta}d\theta=0
\end{displaymath}
for all $\phi_+\in\defD(p_+)$, and thus for $Q_+\psi_+$ to belong to $\defD(Q_+^*)$ it is necessary and sufficient that

\begin{displaymath}
\lim_{\epsilon\to0}\epsilon\int_0^{2\pi}\left(\dzbar(e^{-h}\psi_+)e^h\right)\Big|_{z=\epsilon e^{i\theta}}\overline{\phi_+(\epsilon e^{i\theta})}e^{-i\theta}d\theta=0
\end{displaymath}
for all $\phi_+\in\defD(p_+)$. We know that $\psi_+$ has asymptotics $\psi_+\sim c_{-\alpha}^+r^{-\alpha}+c_{\alpha}^+r^{\alpha}+c_{\alpha-1}^+r^{\alpha-1}e^{-i\theta}+c_{1-\alpha}^+r^{1-\alpha}e^{-i\theta}+O(r^\gamma)$ and that$\dzbar=\frac{e^{i\theta}}{2}\left(\frac{\partial}{\partial r}+\frac{i}{r}\frac{\partial}{\partial\theta}\right)$ in polar coordinates. A calculation gives

\begin{displaymath}
\epsilon\dzbar(e^{-h}\psi_+)e^he^{-i\theta}\Big|_{z=\epsilon e^{i\theta}}\sim -2\alpha c_{-\alpha}^+\epsilon^{-\alpha}+2(1-\alpha)c_{1-\alpha}^+\epsilon^{1-\alpha}e^{-i\theta}+O(r^\gamma),
\end{displaymath}
hence we must have

\begin{equation}
\label{eq:spinupkrav}
\lim_{\epsilon\to0}
\int_0^{2\pi}\left(-2\alpha c_{-\alpha}^+\epsilon^{-\alpha}+2(1-\alpha)c_{1-\alpha}^+\epsilon^{1-\alpha}e^{-i\theta}\right)\overline{\phi_+(\epsilon e^{i\theta})}d\theta=0
\end{equation}
for all $\phi_+\in\defD(p_+)$. A similar calculation for the spin-down component yields
\begin{equation}
\label{eq:spindownkrav}
\lim_{\epsilon\to0}
\int_0^{2\pi}\left(2\alpha c_{\alpha}^-\epsilon^{\alpha}+2(\alpha-1)c_{\alpha-1}^-\epsilon^{\alpha-1}e^{i\theta}\right)\overline{\phi_-(\epsilon e^{i\theta})}d\theta=0.
\end{equation}
We will now calculate what parameters $\nu_0^\pm$ and $\nu_1^\pm$ the Maximal and EV Pauli extensions correspond to. To do so, it is enough to study the asymptoics of the functions in the form core.

Let us first consider the Maximal Pauli extension. Functions on the form $(\phi_0^+c/z)e^h$ constitute a form core for $\pfm^h_+$, where $\phi_0$ is smooth. Hence there are elements in $\defD(\pfm^h_+)$ that asymptotically behave as $r^\alpha$ and also elements with asymptotics $r^{\alpha-1}e^{-i\theta}$. According to~\eqref{eq:spinupkrav} this means that $c_{-\alpha}^+$ and $c_{1-\alpha}^+$ must be zero. Similarly, the elements that behave like $r^{-\alpha}$ and elements that behave like $r^{1-\alpha}e^{i\theta}$ constitute a form core for $\pfm^h_-$, which by~\eqref{eq:spindownkrav} forces $c_{\alpha}^-$ and $c_{\alpha-1}^-$ to be zero. The parameters $\nu_0^\pm$ and $\nu_1^\pm$ are given by $\nu_0^+=c_{\alpha}^+/c_{-\alpha}^+=\infty$, $\nu_1^+=c_{1-\alpha}^+/c_{\alpha-1}^+=0$, $\nu_0^-=c_{\alpha}^-/c_{-\alpha}^-=0$ and $\nu_1^-=c_{1-\alpha}^-/c_{\alpha-1}^-=\infty$. Hence the Maximal Pauli operator $\paulim_h$ can be approximated in the sense of~\cite{bopu}.

Let us now consider the EV Pauli extension, and study the case when $\alpha\in(0,1/2)$. The case $\alpha<0$ follows in a a similar way. A form core for $\pfe^h_+$ is given by $e^h\phi_0$ where $\phi_0$ is smooth, see~\cite{ervo}. These functions have asymptotic behavior $r^\alpha$. From~\eqref{eq:spinupkrav} follows that $c_{-\alpha}^+$ must vanish. However, $\psi_+$ belonging to $\defD(Q_+)$ must also belong to $\defD(\pfe^h_+)$ and since the functions in the form core for $\pfe^h_+$ behave as $r^{\alpha}$ or nicer, we see that the term $c_{\alpha-1}^+r^{\alpha-1}e^{-i\theta}$ gets too singular to be in $\defD(Q_+)$ if $c_{\alpha-1}^+\neq 0$, and hence $c_{\alpha-1}^+$ must be zero.

Similarly, a form core for $\pfe^h_-$ is given by $e^{-h}\phi_0$, with $\phi_0$ smooth. Functions in this form core have asymptotic behavior $r^{-\alpha}$ or $r^{-\alpha+1}e^{i\theta}$ which forces $c_{\alpha}^-$ and $c_{\alpha-1}^-$ to be zero.

Hence the parameters $\nu_0^\pm$ and $\nu_1^\pm$ are given by $\nu_0^+=c_{\alpha}^+/c_{-\alpha}^+=\infty$, $\nu_1^+=c_{1-\alpha}^+/c_{\alpha-1}^+=\infty$, $\nu_0^-=c_{\alpha}^-/c_{-\alpha}^-=0$ and $\nu_1^-=c_{1-\alpha}^-/c_{\alpha-1}^-=\infty$.

We conclude that the spin-up part of $\paulie_h$ can not be approximated in the sense of \cite{bopu}, while the spin-down part can.

\section*{Acknowledgements}
\noindent I would like to thank my supervisor Professor Grigori Rozenblum for introducing me to this problem and for giving me all the support I needed.

\end{document}